\documentclass{article}
\usepackage{ijcai24}

\usepackage{times}
\usepackage{soul}
\usepackage{url}
\usepackage[hidelinks]{hyperref}
\usepackage[utf8]{inputenc}
\usepackage[small]{caption}
\usepackage{graphicx}
\usepackage{amsmath}
\usepackage{amsthm}
\usepackage{booktabs}
\usepackage{algorithm}
\usepackage{algorithmic}
\usepackage[switch]{lineno}

\usepackage{multirow}
\usepackage{xcolor}
\usepackage{amssymb}
\usepackage{tikz}
\usepackage[edges]{forest}
\usepackage{mydef}

\title{Generative AI for Controllable Protein Sequence Design: A Survey}

\author{
    Author Name
    \affiliations
    Affiliation
    \emails
    email@example.com
}

\author{
Yiheng Zhu$^1$
\and
Zitai Kong$^1$\and
Jialu Wu$^1$\and
Weize Liu$^1$\and
Yuqiang Han$^2$\and
Mingze Yin$^1$\and
Hongxia Xu$^1$\and
Chang-Yu Hsieh$^1$\And
Tingjun Hou$^1$\\
\affiliations
$^1$Zhejiang University\\
$^2$The Chinese University of Hong Kong\\
\emails
\{zhuyiheng2020, kongzitai, jialuwu, weizeliu, 22218878, Einstein, kimhsieh, tingjunhou\}@zju.edu.cn,
yuqianghan@cuhk.edu.hk
}

\begin{document}

\maketitle

\begin{abstract}
    The design of novel protein sequences with targeted functionalities underpins a central theme in protein engineering, impacting diverse fields such as drug discovery and enzymatic engineering. However, navigating this vast combinatorial search space remains a severe challenge due to time and financial constraints. This scenario is rapidly evolving as the transformative advancements in AI, particularly in the realm of generative models and optimization algorithms, have been propelling the protein design field towards an unprecedented revolution. In this survey, we systematically review recent advances in generative AI for controllable protein sequence design. To set the stage, we first outline the foundational tasks in protein sequence design in terms of the constraints involved and present key generative models and optimization algorithms. We then offer in-depth reviews of each design task and discuss the pertinent applications. Finally, we identify the unresolved challenges and highlight research opportunities that merit deeper exploration.
\end{abstract}

\section{Introduction}
Proteins are indispensable biomolecules that execute the myriad of biological processes fundamental to life, such as catalyzing enzymatic reactions and mediating immune responses. Motivated by the versatility exhibited by natural proteins, designing novel amino acid sequences that encode proteins with desired functions has been a central challenge in bioengineering. Nevertheless, due to the vast exploration space of possible proteins, conventional experimental methods, such as directed evolution~\cite{arnold1998design}, are prohibitively time-intensive and money-consuming. To overcome these challenges, recent years have witnessed the trend of leveraging generative AI technologies to explore the biochemical space intelligently. Fueled by advances in generative AI, the field of protein design is experiencing an unprecedented revolution.

Evolutionary-scale protein sequence databases, as opposed to limited and static structural data that yield a biased subset of the protein design space, lay the groundwork for sequence-based protein design. Deep generative models offer a promising tool for capturing the distribution underlying natural protein sequences to generate novel and diverse proteins. Moreover, making protein design controllable is crucial for the transition from theory to practice. Ideally, a practical method ought to generate proteins that adhere to task-specific structural and functional constraints. To this end, researchers have utilized conditional generative models and optimization algorithms to realize such tailored protein design.

\begin{figure*}[t]
\centering
\includegraphics[width=0.92\textwidth]{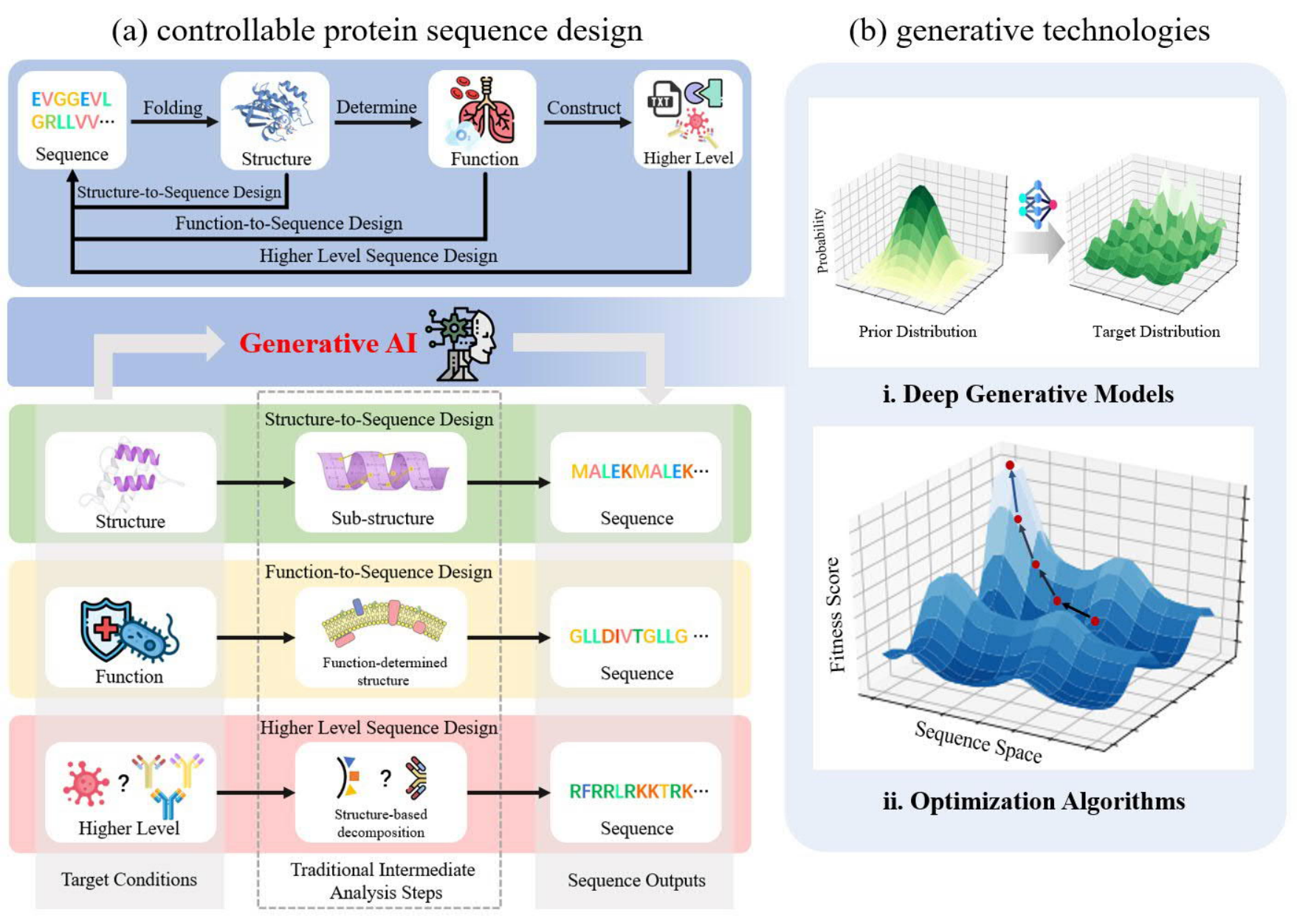} 
\caption{Illustration for controllable protein sequence design. (a) The upper diagram delineates the four integral tiers of the protein central dogma. Amino acid sequences fold to form specific protein structures, which determine protein functions. These varied functions integrate to perform higher-level actions. Then protein sequence design tasks can be bifurcated into structure-to-sequence, function-to-sequence, and higher-level sequence design. The lower diagram highlights that generative AI empowers designers to sidestep intricate intermediate steps associated with traditional methods, facilitating protein design in an end-to-end manner. (b) Related generative AI technologies primarily incorporate (i) deep generative models and (ii) optimization methods (see Section~\ref{sec:preliminaries} for more details).}
\label{fig:framework}
\end{figure*}

Protein sequence design stands as a rapidly growing field of scientific discovery, necessitating a thorough review of the extensive literature to sketch the contours of the field. Despite certain surveys detailing advancements~\cite{wu2021protein} and highlighting controllability~\cite{ferruz2022controllable}, they may fall short in igniting enthusiasm for controllable protein sequence design among the machine learning community due to two main reasons: 1) the majority are framed from a biochemical standpoint, potentially inaccessible for researchers without deep domain expertise; 2) they often focus on selected aspects of the literature in terms of methodologies and tasks. For instance, compared to \textit{de novo} design, the equally important theme of mutation-based optimization tends to be overlooked. In contrast, our survey endeavors to bridge this gap. We seek to demystify protein sequence design for machine learning researchers by systematically categorizing various controllable design tasks according to the constraints involved. We hope this survey reaches both the machine learning and biochemistry fields and inspires further collaboration towards developing more sophisticated and controllable algorithms.

\tikzstyle{leaf}=[draw=hiddendraw,
    rounded corners,minimum height=1em,
    fill=hidden-orange!40,text opacity=1, align=center,
    fill opacity=.5,  text=black,align=left,font=\scriptsize,
    inner xsep=3pt,
    inner ysep=1pt,
    ]
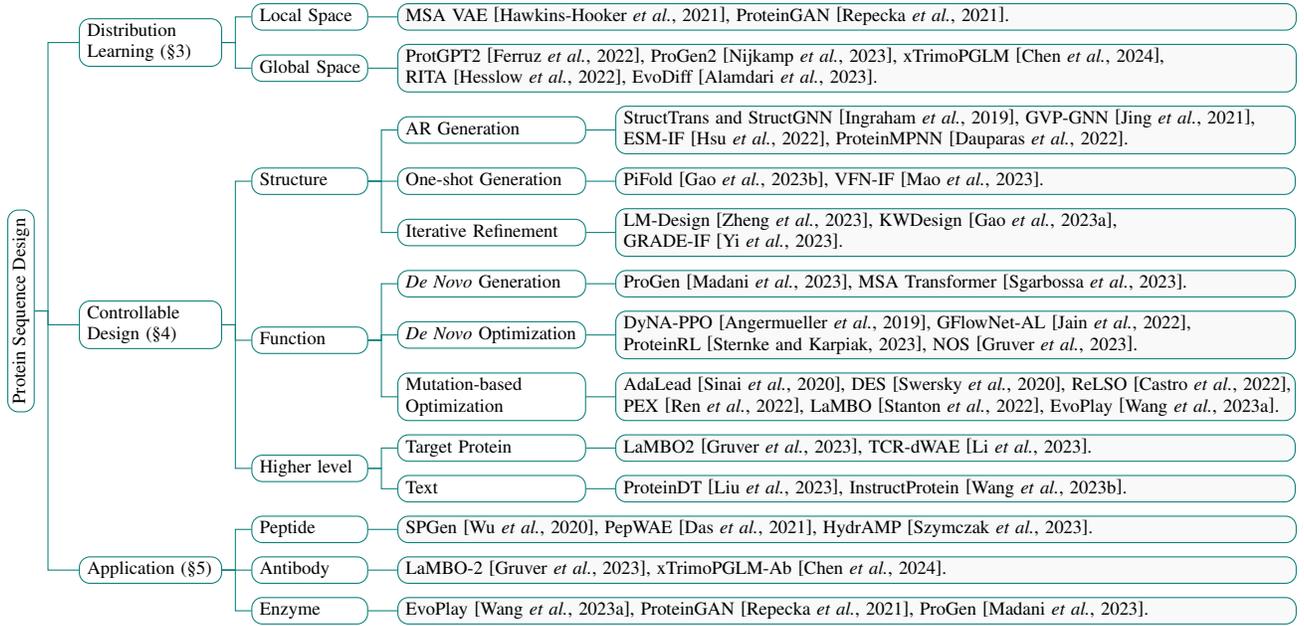
\begin{figure*}[ht]
\centering
\begin{forest}
  for tree={
  forked edges,
  grow=east,
  reversed=true,
  anchor=base west,
  parent anchor=east,
  child anchor=west,
  base=middle,
  font=\scriptsize,
  rectangle,
  draw=hiddendraw,
  rounded corners,align=left,
  minimum width=2em,
    s sep=5pt,
    inner xsep=3pt,
    inner ysep=1pt,
  },
  where level=1{text width=4.5em}{},
  where level=2{text width=6em,font=\scriptsize}{},
  where level=3{font=\scriptsize}{},
  where level=4{font=\scriptsize}{},
  where level=5{font=\scriptsize}{},
  [Protein Sequence Design,rotate=90,anchor=north,edge=hiddendraw
    [Distribution\\Learning (\S\ref{sec:distribution}), edge=hiddendraw,text width=4.78em
     [Local Space, text width=3.78em, edge=hiddendraw
        [
        MSA VAE~\cite{hawkins2021generating}{,} 
        ProteinGAN~\cite{repecka2021expanding}{.}
        ,leaf,text width=33.38em, edge=hiddendraw
        ]
     ]
    [Global Space, text width=3.78em, edge=hiddendraw
        [ProtGPT2~\cite{ferruz2022protgpt2}{,}
        ProGen2~\cite{nijkamp2023progen2}{,}
        xTrimoPGLM~\cite{chen2024xtrimopglm}{,} \\
        RITA~\cite{hesslow2022rita}{,}
        EvoDiff~\cite{alamdari2023protein}{.}
        ,leaf,text width=33.38em, edge=hiddendraw]
    ]
    ]
    [Controllable\\Design (\S\ref{sec:control}), edge=hiddendraw,text width=4.78em
      [Structure, text width=3.78em, edge=hiddendraw
        [AR Generation, edge=hiddendraw, text width=6.5em
            [StructTrans and StructGNN~\cite{ingraham2019generative}{,} 
            GVP-GNN~\cite{jing2021learning}{,} \\
            ESM-IF~\cite{hsu2022learning}{,}
            ProteinMPNN~\cite{dauparas2022robust}{.}
            ,leaf,text width=25.1em, edge=hiddendraw]
        ]
        [One-shot Generation, edge=hiddendraw, text width=6.5em
            [PiFold~\cite{gao2023pifold}{,}
            VFN-IF~\cite{mao2023novo}{.}
            ,leaf,text width=25.1em, edge=hiddendraw]
        ]
        [Iterative Refinement, edge=hiddendraw, text width=6.5em
            [
                LM-Design~\cite{zheng2023lm_design}{,}
                KWDesign~\cite{gao2023knowledge}{,} \\
                GRADE-IF~\cite{yi2023graph}{.}
                ,leaf,text width=25.1em, edge=hiddendraw
            ]
        ]
      ]
    [Function, text width=3.78em, edge=hiddendraw
      [\textit{De Novo} Generation, text width=6.5em, edge=hiddendraw
        [
            ProGen~\cite{madani2023large}{,}
            MSA Transformer~\cite{sgarbossa2023generative}{.}
        ,leaf,text width=25.1em, edge=hiddendraw]
      ]
      [\textit{De Novo} Optimization, text width=6.5em, edge=hiddendraw
        [
            DyNA-PPO~\cite{angermueller2019model}{,}
            GFlowNet-AL~\cite{jain2022biological}{,} \\
            ProteinRL~\cite{sternke2023proteinrl}{,}
            NOS~\cite{gruver2023protein}{.}
        ,leaf,text width=25.1em, edge=hiddendraw]
      ]
      [Mutation-based \\ Optimization, text width=6.5em, edge=hiddendraw
        [
            AdaLead~\cite{sinai2020adalead}{,}
            DES~\cite{swersky2020amortized}{,} 
            ReLSO~\cite{castro2022transformer}{,} \\
            PEX~\cite{ren2022proximal}{,}
            LaMBO~\cite{stanton2022accelerating}{,}
            EvoPlay~\cite{wang2023self}{.}
        ,leaf,text width=25.1em, edge=hiddendraw]
      ]
    ]
    [Higher level, text width=3.78em, edge=hiddendraw
        [Target Protein, text width=6.5em, edge=hiddendraw
        [
            LaMBO2~\cite{gruver2023protein}{,} 
            TCR-dWAE~\cite{li2023disentangled}{.}
        ,leaf,text width=25.1em, edge=hiddendraw
        ]
        ]
        [Text, text width=6.5em, edge=hiddendraw
            [
                ProteinDT~\cite{liu2023text}{,}
                InstructProtein~\cite{wang2023instructprotein}{.}
            ,leaf,text width=25.1em, edge=hiddendraw
            ]
        ]
    ]
    ]
    [Application (\S\ref{sec:app}), edge=hiddendraw,text width=4.78em
        [Peptide, text width=3.78em, edge=hiddendraw
            [
                SPGen~\cite{wu2020signal}{,}
                PepWAE~\cite{das2021accelerated}{,}
                HydrAMP~\cite{szymczak2023discovering}{.}
            ,leaf,text width=33.38em, edge=hiddendraw
            ]
         ]
         [Antibody, text width=3.78em, edge=hiddendraw
            [
                LaMBO-2~\cite{gruver2023protein}{,}
                xTrimoPGLM-Ab~\cite{chen2024xtrimopglm}{.}
            ,leaf,text width=33.38em, edge=hiddendraw
            ]
         ]
        [Enzyme, text width=3.78em, edge=hiddendraw
            [
                EvoPlay~\cite{wang2023self}{,}
                ProteinGAN~\cite{repecka2021expanding}{,}
                ProGen~\cite{madani2023large}{.} 
            ,leaf,text width=33.38em, edge=hiddendraw]
        ]
    ]
  ]
\end{forest}
\caption{A taxonomy of protein sequence design methods with representative examples.}
\label{fig:texonomy}
\end{figure*}

The remainder of this paper is organized as follows: We first outline the foundational tasks involving controllable protein sequence design, along with relevant generative models and optimization algorithms (Section~\ref{sec:preliminaries}). We then encapsulate the prevalent paradigms for distribution learning on natural sequence space (Section~\ref{sec:distribution}). Next, we provide an extensive review of each controllable design task (Section~\ref{sec:control}). We also examine the pertinent applications (Section~\ref{sec:app}). Finally, we conclude with the open challenges and point to potential opportunities for further exploration (Section~\ref{sec:future}). A detailed breakdown is depicted in Figure~\ref{fig:texonomy}. 

\section{Preliminaries}
\label{sec:preliminaries}

\subsection{Problem Formulation}
Generally, a protein can be represented as a sequence of amino acids $x=[x_1, x_2, \dots, x_L]$ of length $L$, where each amino acid $x_i$ is drawn from a vocabulary $\mathcal{V}$ consisting of 20 standard amino acids. Controllable protein sequence design aims to generate plausible proteins that satisfy some desired constraints $c$, which may vary across different tasks. Formally, we pose the controllable protein sequence design as the problem of sampling $x$ from $p(x|c)$. 

A fundamental tenet of molecular biology is that ``sequence determines structure, structure determines function''~\cite{branden2012introduction}. Consequently, we divide the meaningful control conditions into three categories, i.e., structural, functional, and higher-level constraints, as illustrated in Figure~\ref{fig:framework}.

\subsection{Deep Generative Models}
Herein, we concisely outline the core principles behind representative generative frameworks.

\textbf{Autoregressive models (ARMs)}~\cite{bengio2000neural} learn the high-dimensional joint distribution $p(x)$ by factorizing it into the product of univariate conditionals as follows:
\begin{equation}
    p(x)=\prod_{i=1}^{L} p(x_i|x_1,x_2,\dots,x_{i-1})
\end{equation} 
ARMs require a predefined order in which to sequentially generate $x$. While the choice of ordering can be clear for some modalities (e.g., text and audio), it is not obvious for others such as images and graphs.

\textbf{Variational Autoencoders (VAEs)}~\cite{kingma2013auto} optimize the evidence lower bound (ELBO) of the intractable likelihood $p(x)$. VAEs introduce a latent variable $z$ to derive the ELBO: 
\begin{equation}
    \resizebox{.91\linewidth}{!}{$
                \displaystyle
    \begin{aligned}
    \log{p(x)}
    & = \log{\int p(z) p(x|z) \mathrm{d}z}\\
    & \ge \mathbb{E}_{q_{\phi}(z|x)} \big[ \log{p(x|z)} \big] - D_{KL}(q_{\phi}(z|x) || p(z))\\
    & \triangleq \text{ELBO}
    \end{aligned}
    $}
\end{equation}
where $p(z)$ is the prior distribution, $q_{\phi}(z|x)$ is the amortized variational posterior. The first term of the ELBO is referred to as the (negative) reconstruction error, while the second term is the KL-divergence which quantifies the discrepancy between the learned distribution $q_{\phi}(z|x)$ and prior distribution $p(z)$.

\textbf{Generative Adversarial Networks (GANs)}~\cite{goodfellow2020generative} are a class of likelihood-free models, aiming to generate vivid fake data indistinguishable from real data. GANs are composed of a discriminator $D$ that tries to distinguish the generated data from real data, and a generator $G$ that aims to generate data convincing enough to deceive the discriminator. This objective is achieved through adversarial training:
\begin{equation} 
\begin{aligned}
& \min_{G} \max_{D} \mathcal{L}(D, G) \\
= & \, \mathbb{E}_{x \sim p_{data} } [\log D(x)] + \mathbb{E}_{z \sim p(z)} [\log (1-D(G(z))]
\end{aligned}
\end{equation}

\textbf{Diffusion models}~\cite{sohl2015deep,ho2020denoising} are latent variable models inspired by non-equilibrium thermodynamics. The essential idea is to define a forward diffusion process that gradually corrupts the data into pure noise and learn the reverse process to reconstruct the data from noise distributions. 

Most works in continuous spaces define the forward process as $q(x_t|x_{t-1})={\mathcal N}(x_t|\sqrt{1-\beta_t}x_{t-1},\beta_t{\mathbb I})$ to corrupts the data $x_0 \sim p(x)$ into a sequence of increasingly noisy latent variables $x_{1:T} = x_1, x_2, \dots, x_T$. The reverse process can then be computed as $q(x_{t-1}|x_t, x_0)=\mathcal N(x_{t-1}|{\widetilde{\mu}}_t(x_t,x_0), {\widetilde{\beta}}_t\mathbb I)$ when conditioned on $x_0$ using Bayes theorem. Let $\alpha_t=1-\beta_t$ and $\hat{\alpha}_t=\prod_{s=1}^{t}{\alpha_t}$, the parameters $\widetilde{\mu_t}(x_t,x_0)=\frac{\sqrt{{\hat{\alpha}}_{t-1}}\beta_t}{1-\hat{\alpha}_t}x_0+\frac{\sqrt{\alpha_t}(1-\hat{\alpha}_{t-1})}{1-\hat{\alpha}_t}x_t$ and ${\widetilde{\beta}}_t=\frac{1-\hat{\alpha}_{t-1}}{1-\hat{\alpha}_t}\beta_t$ are analytical. The objective of diffusion models is derived from the variational lower bound of $p(x)$, delineated as follows:
\begin{equation} 
    \resizebox{.91\linewidth}{!}{$
            \displaystyle    
    \begin{aligned}
    & \mathcal{L}_{\text{VLB}} 
     = \underbrace{KL[q(x_T|x_0) || p_\theta(x_T)]}_{\mathcal{L}_T} \\
    & ~~~ + \sum_{t=2}^T \underbrace{KL[q(x_{t-1}|x_t,x_0) || p_\theta(x_{t-1}|x_t)]}_{\mathcal{L}_{t-1}} \underbrace{- \mathbb{E}_q [\log p_\theta(x_0|x_1)]}_{\mathcal{L}_0}
    \end{aligned}
    $}
\end{equation}
where $\mathcal{L}_T$ is a constant and $\mathcal{L}_0$ can be estimated using a separate model. Diffusion models commonly adopt a neural network ${\epsilon}_{\theta}$ to predict the noise such that $\{\mathcal{L}_{t-1}\}_{t=2}^T$ is reduced to $\mathbb{E}_{x_0,t} \Big[\| \epsilon_t - \epsilon_\theta(\sqrt{\bar \alpha_t}x_0 + \sqrt{1-\bar \alpha_t}\epsilon_t, t) \|^2 \Big]$.

While diffusion models with Gaussian perturbation excel in generating continuous signals, their direct application to discrete data generation is not feasible. To bridge this gap, researchers have explored two main strategies for adapting diffusion models to discrete domains: the first embeds discrete symbols into a continuous space~\cite{li2022diffusion}, and the second defines a diffusion process that inherently operates within discrete space~\cite{austin2021structured}.

\subsection{Optimization algorithms}
We overview the widely-used optimization algorithms designed to optimize the objective function or reward.

\textbf{Latent space optimization (LSO)}~\cite{gomez2018automatic} circumvents the discrete search space by performing optimization over the low-dimensional continuous latent space learned by generative models. Following the navigation of the latent space, the optimized latent representations are then decoded back into sequences.

\textbf{Reinforcement learning (RL)}~\cite{sutton2018reinforcement} seeks to learn an optimal policy that maximizes the reward. The RL problem is typically formalized within the Markov Decision Process (MDP) framework, which consists of states, actions, transitions, and a reward function. In RL, agents investigate their environment via a process of trial and error, learning to take actions that lead to the highest rewards. Recently, the RL-like algorithm GFlowNet~\cite{bengio2021flow} has emerged, offering
a method to learn a stochastic policy that sequentially constructs objects with probability proportional to reward, carving a niche for itself in scientific discovery~\cite{jain2022biological,zhu2023sample}.

\textbf{Evolution algorithms} are popular heuristic algorithms inspired by natural evolutionary processes. Beginning with a population of individuals, where each represents a potential solution within the search space of a given problem, evolution algorithms mimic the stochastic processes of mutation and/or recombination to facilitate exploration within the search space. Individuals with higher fitness values are preferentially selected in this adaptive environment.

\textbf{Bayesian optimization (BO)}~\cite{shahriari2015taking} provides a sample-efficient active learning framework for the global optimization of expensive black-box functions. During each iteration, the principal strategy is to build a probabilistic surrogate model that approximates the actual black-box functions on the observed data and then optimize an acquisition function, built upon the surrogate model, to identify informative candidates which offer the greatest utility for the next cycle of evaluations.

\section{Distribution Learning of Natural Sequences}
\label{sec:distribution}
Comprehending the underlying data distributions of natural protein sequences is crucial for exploring the protein landscape. Deep generative models serve as powerful tools for learning these distributions and generating novel sequences that extend beyond the currently explored regions of the protein sequence space. We delineate the models specifically developed to characterize either the local or global protein space.

\subsection{Local Sequence Space}
\label{subsec:local}
Early works train generative models on multiple sequence alignment (MSA) of a specific family of interest to capture statistical patterns of amino-acid relationships and generate novel proteins belonging to that family. The Potts model is a well-defined statistical energy model for capturing the first- and second-order evolutionary dependencies. Representatively, the Potts model fitted with direct coupling analysis is used to design chorismate mutase enzymes~\cite{russ2020evolution}. Beyond conventional statistical approaches, various deep generative models have also been reasonably applied to capture higher-order dependencies. Successful examples include \textbf{MSA VAE}~\cite{hawkins2021generating}, and \textbf{ProteinGAN}~\cite{repecka2021expanding}.

\subsection{Global Sequence Space}

\paragraph{Protein language models.}
The advent of autoregressive large language models (LLMs) such as GPT-3~\cite{brown2020language}, which can generate text with human-like capabilities, has opened new avenues in protein sequence design due to the structural parallels between natural language and protein sequences—both are sequential systems of fundamental units. Building on the flexible and powerful Transformer architecture~\cite{vaswani2017attention}, renowned for its proficiency in capturing long-range dependencies within sequences, protein language models (PLMs) pre-trained on a large-scale corpus of protein sequences succeed in capturing intrinsic sequence features and generating novel protein sequences that exhibit properties analogous to those of natural proteins. \textbf{ProtGPT2}~\cite{ferruz2022protgpt2} is an autoregressive PLM with 738 million parameters, showcasing the GPT-2 architecture's capabilities in generating novel protein sequences \textit{de novo}. It utilizes the BPE algorithm for tokenizing sequences rather than directly employing amino acids as tokens. Having been pre-trained on nearly 50 million protein sequences, ProtGPT2 is capable not only of generating sequences that conform to the patterns of natural proteins but also of sampling unexplored regions of protein space. \textbf{ProGen2}~\cite{nijkamp2023progen2} models are based on the GPT-J architecture and have been scaled up to 6.4 billion parameters, exhibiting enhanced proficiency in capturing the distribution of observed evolutionary sequences and making zero-shot fitness predictions. \textbf{xTrimoPGLM}~\cite{chen2024xtrimopglm} adopts the backbone of the general language model and utilizes in-place token prediction alongside autoregressive blank infilling as training objectives. This innovative pre-training framework enables xTrimoPGLM to perform both \textit{de novo} design and targeted sequence infilling. Scaling laws for LLMs~\cite{kaplan2020scaling} delineate the correlation between model performance and size, thereby providing the theoretical basis for the strategic enlargement of these models. Hesslow et al.~\shortcite{hesslow2022rita} present \textbf{RITA}, a suite of generative PLMs with up to 1.2 billion parameters, and conduct the first systematic analysis of how capabilities evolve relative to model size within the protein domain, revealing benefits associated with scaling up. More recently, \textbf{xTrimoPGLM-100B} significantly outperforms other advanced PLMs across both protein understanding and generation tasks, further emphasizing scaling effects.

\paragraph{Diffusion models.}
The emergence of diffusion models as dominant frameworks in image synthesis has attracted considerable interest in their application to protein design, offering distinct advantages over left-to-right ARMs, including flexible iterative refinement enabled by order-agnostic decoding and the trade-off between sample quality and inference efficiency. \textbf{EvoDiff}~\cite{alamdari2023protein} is the first foundation diffusion model for protein design trained on evolutionary-scale protein sequence data. EvoDiff presents two variants, EvoDiff-Seq and EvoDiff-MSA, which are grounded in two famous discrete diffusion frameworks: D3PM~\cite{austin2021structured} and OADM~\cite{hoogeboom2022autoregressive}. Extensive experiments show that EvoDiff can generate diverse and structurally plausible proteins encompassing the breadth of natural sequence and functional spaces. Owing to the universal applicability of sequence-based design paradigm, EvoDiff can produce proteins inaccessible to structure-based models, especially those with disordered regions, while preserving the capacity to engineer scaffolds for functional structural motifs.

\subsection{Databases}
\textbf{UniProt}~\cite{uniprot2019uniprot} serves as an extensive protein repository, documenting hundreds of millions of proteins spanning all branches of life. It offers three unique databases: UniProtKB, UniRef, and UniParc. Among these, the UniRef databases, which cluster sequence sets at various levels of sequence identity, are broadly utilized in pre-training PLMs. Besides, the \textbf{OAS} database~\cite{olsen2022observed} is specifically designed to curate and annotate immune repertoires suitable for extensive analytical research. It furnishes comprehensive nucleotide and amino acid profiles for each antibody, enriched with additional sequence annotations. OAS features a vast collection of more than 2.4 billion unpaired and over 1.5 million paired antibody sequences.

\section{Controllable Protein Sequence Design}
\label{sec:control}
\subsection{Structure-to-sequence Design}
Structure-to-sequence design, which is often referred to as the protein inverse folding problem, aims to automatically identify the protein sequences that fold into the given backbone structures. Formally, the inverse folding problem is represented as learning the conditional distribution $p(x|r)$, where $r=[r_1, r_2, \dots, r_n]\in \mathbb{R}^{n\times 4\times 3}$ denotes the 3D coordinates of backbone atoms in each residue, namely N, $\text{C}_\alpha$, C, and O.

Typically, Protein inverse folding models consist of an encoder that represents the protein backbone and a decoder tasked with generating the protein sequence. The backbone structure is commonly represented as a proximity graph over amino acid nodes. According to the generative manner, modern inverse folding models are categorized into autoregressive, one-shot, and iterative models.

\paragraph{Autoregressive generation.}  
Ingraham et al.~\shortcite{ingraham2019generative} firstly framed inverse folding as a structure-to-sequence learning problem, proposing to predict protein sequences from N to C terminus in an autoregressive manner using backbone features extracted by the structure encoder. They define a local coordinate system at each amino acid node to derive relative distance, direction, and orientation for edge features, and compute the backbone dihedral angles to constitute node features. Further, they proposed two variants, \textbf{StructTrans} and \textbf{StructGNN}, to serve as the structure encoders. Following Ingraham et al.~\shortcite{ingraham2019generative}, a succession of studies has adopted the autoregressive framework, achieving further enhancement in performance. For instance, \textbf{GVP-GNN}~\cite{jing2021learning} replaces the standard multi-layer perceptrons in GNNs with geometric vector perceptrons, which can effectively operate on both scalar and geometric features. \textbf{ESM-IF}~\cite{hsu2022learning} augments training data by incorporating 12M structures predicted by AlphaFold2, and examines three model architectures, namely GVP-GNN, GVP-GNN-large, and GVP-Transformer. Building on the foundation of StructGNN, \textbf{ProteinMPNN}~\cite{dauparas2022robust} verifies that incorporating distances between N, $\text{C}_\alpha$, C, O, and a virtual $\text{C}_\beta$ as extra input features and employing edge updates in the encoder can lead to a marked enhancement of performance. To facilitate a wide array of applications, ProteinMPNN adopts an order-agnostic ARM where a random decoding order is sampled. 

\paragraph{One-shot generation.}
Given that the autoregressive models tend to have low inference speed, some researchers have investigated one-shot methods that facilitate parallel generation of multiple tokens. \textbf{PiFold}~\cite{gao2023pifold} replaces the autoregressive decoder with stacked PiGNNs and improves the quality and efficiency simultaneously, providing the first evidence that the expressive encoder could play a more crucial role than the autoregressive decoder in inverse folding. To be precise, PiFold integrates a broader range of features, along with learnable virtual atoms designed to capture information not accounted for by real atoms, and employs PiGNN layers to effectively learn from multi-scale residue interactions by accounting for feature dependencies at the node, edge, and global levels. Advancing in this direction, \textbf{VFN-IF}~\cite{mao2023novo} introduces a more expressive encoder called Vector Field Networks, which extract hierarchical residual representations at both the real atom and frame-anchored virtual atom levels, thereby surpassing PiFold.

\paragraph{Iterative refinement.}
Another promising approach involves generating an initial sequence then iteratively refining the generated sequence. Motivated by the insight that PLMs can learn evolutionary knowledge from the universe of natural protein sequences, \textbf{LM-Design}~\cite{zheng2023lm_design} reprograms sequence-based PLMs, such as ESM-1b~\cite{rives2021biological}, for inverse folding. By integrating a lightweight structural adapter, LM-Design equips PLMs with structural comprehension. LM-Design undertakes training through conditional masked language modeling, which allows it to reconstruct a native sequence from a corrupted version, facilitating the iterative refinement of the predicted sequence. \textbf{KWDesign}~\cite{gao2023knowledge} also capitalizes on the common protein knowledge extracted from PLMs to refine low-quality residues. Specifically, KWDesign introduces a confidence-aware knowledge-tuning module that fuses multimodal pre-trained knowledge to generate more rational sequences. \textbf{GRADE-IF}~\cite{yi2023graph} controls the intermediate refinement process by applying the principles of discrete diffusion diffusion probabilistic models (D3PM)~\cite{austin2021structured}. To constrain the generative space to a reduced subspace reflecting evolutionary pressures, the Blocks Substitution Matrix (BLOSUM) is employed as the transition matrix to guide the forward diffusion process. A primary benefit of GRADE-IF is its capacity to encompass diverse plausible solutions.

\subsubsection{Datasets and Benchmarks}
Conventionally, inverse folding models are benchmarked using the CATH4.2 and CATH 4.3 datasets~\cite{orengo1997cath}, with perplexity and recovery rate serving as the key evaluation metrics. Additionally, the TS50 and TS500 datasets~\cite{li2014direct} are commonly used as independent test sets for evaluating the generalization ability. Advanced challenges encompass the design of multi-chain complexes, antibodies, and \textit{de novo} proteins~\cite{zheng2023lm_design}. Recently, Gao et al.~\shortcite{gao2023proteininvbench} established ProteinInvBench, the first comprehensive benchmark for inverse folding, aiming at extending state-of-the-art models to new tasks and evaluating them based on a variety of metrics such as confidence, diversity, and sc-TM.

\subsection{Function-to-sequence Design}
The goal of function-to-sequence design is to generate novel, diverse, and biologically plausible proteins with desired functions. Based on the manner in which the objective is achieved, we roughly divided existing methods into three distinct categories: \textit{De novo} generation, \textit{De novo} optimization, and mutation-based optimization. Note that most optimization methods can be adopted as the acquisition function optimizer in BO, enabling online model-based optimization. For the sake of conciseness, we will focus solely on the optimization mechanisms of these methods without delving into the specifics of their application in BO.

\subsubsection{\textit{De novo} generation}
The most direct approach is to train generative models on known active proteins. The family-specific generative models discussed in Section~\ref{subsec:local} exemplify this methodology. Despite that they are capable of generating functional proteins, their generalizability to sequence space outside the scope of the training family is limited. Consequently, later works tend to train a unified conditional model across diverse protein families. Inspired by CTRL~\cite{keskar2019ctrl}, \textbf{ProGen}~\cite{madani2023large} designs a decoder-only transformer for autoregressive generation and incorporates control tags (e.g., protein family, biological process, and molecular function), which are metadata prepended to the protein during training, to enable controllable protein design. \textbf{MSA Transformer}~\cite{rao2021msa} pre-trained with masked language modeling, has been leveraged to generate synthetic MSAs by performing an iterative mask-then-fill procedure on the input natural MSAs~\cite{sgarbossa2023generative}.

\subsubsection{\textit{De novo} optimization}
Another promising direction is to guide an unconditional generative model with a discriminative surrogate model to produce protein sequences that exhibit elevated levels of functionality, referred to as \textit{fitness}. Given the intensive demands of laboratory experiments, surrogate sequence-function models are commonly built to act as substitutes for wet-lab validation.

RL provides a flexible framework that guides modern deep generative models to navigate the protein fitness landscape. \textbf{DyNA-PPO}~\cite{angermueller2019model} is a pioneer in employing model-based policy optimization and diversity-promoting reward functions for biological sequence design. DyNA-PPO formulates the design problem as an autoregressive MDP, where each action appends an acid amino token to a partially constructed sequence. Likewise, \textbf{GFlownet-AL}~\cite{jain2022biological} strives to learn a stochastic policy that can sequentially generate sequences proportionally to a reward function, proposing a batch of diverse candidates. Fine-tuning the pre-trained PLMs with RL has the potential to enhance performance. Inspired by the achievements of ChatGPT, \textbf{ProteinRL}~\cite{sternke2023proteinrl} has emerged as the first work to design full-length proteins for specific properties by fine-tuning a generative PLM, namely ProGen2.

Diffusion models can be guided to generate continuous samples with the desired conditions, circumventing the necessity for fine-tuning by employing classifier guidance~\cite{dhariwal2021diffusion}. Recent advancements have broadened this strategy to encompass general discrete diffusion models, facilitating guided protein design. \textbf{NOS}~\cite{gruver2023protein} skews the categorical sampling distribution of the diffusion models by exerting gradient guidance upon the continuous representations of discrete sequences, which are expressed in the form of hidden states.

\subsubsection{Mutation-based Optimization}
Contrary to \textit{de novo} optimization, mutation-based optimization focuses on improving a fitness function by modifying an existing wild type, defined as the sequence occurring in nature. In general, the wild type demonstrates decent fitness under the natural selection evolutionary process towards the target function. Traditionally, biologists have approached this problem using a model-free paradigm known as \textit{directed evolution}, which involves iterative cycles of obtaining protein variants by random mutagenesis, followed by screening and selecting high-fitness variants. However, this approach is limited by its reliance on high-throughput functional assays and a greedy hill-climbing search algorithm, which can lead to substantial costs and convergence to local optima. Recent efforts have concentrated on proposing more efficient optimization methods to improve search efficiency.

In classical evolutionary algorithms, only top-fitness sequences are selected and mutated within the evolution cycle, leading to limited diversity. To alleviate this issue, \textbf{AdaLead}~\cite{sinai2020adalead} incorporates recombination of mutations to promote diversity and avoid local optima, establishing itself as both an accessible and robust baseline method. To leverage the natural property of the fitness landscape that a concise set of mutations upon the wild type are usually sufficient to enhance the desired function, \textbf{PEX}~\cite{ren2022proximal} prioritizes the evolutionary search for high-fitness mutants with low mutation counts.

Besides random mutations, an alternative approach is to leverage RL to intelligently guide the search. \textbf{DES}~\cite{swersky2020amortized} employs a convolutional neural network as the policy network to propose specific edits on a target sequence, and these edits are defined as a combination of intended mutation site along with the choice of replacing amino acids. The policy network is subsequently trained using a policy gradient-based RL algorithm. \textbf{EvoPlay}~\cite{wang2023self} also mutates a single-site residue as an action to optimize protein sequences. Likening the protein optimization process to playing pieces on a chessboard, EvoPlay implements an iterative look-ahead Monte Carlo tree search procedure to generate samples for training a policy-value neural network, which in turn is then used to guide the search.

By employing gradient ascent, LSO methods can progress from wild type towards regions of higher fitness. For instance, \textbf{ReLSO}~\cite{castro2022transformer} trains a transformer-based autoencoder with a fitness prediction network jointly, producing a latent space organized by sequence-function relationships. It also introduces norm-based negative sampling and an interpolative sampling penalty to regularize the latent space, further enhancing the optimization process. \textbf{LaMBO}~\cite{stanton2022accelerating} maps discrete sequences to continuous representations using a denoising autoencoder. During the optimization phase, LaMBO follows the gradient in the latent space to reconstruct sequences that have been partially masked. 

\subsection{High-level constraints}
Beyond explicit structural and functional constraints, there exists tremendous knowledge in other formats describing proteins’ higher-level properties. 

\paragraph{Target protein.} 
Target-specific protein design paves the way for the design of functional proteins that possess distinct specificity. A prime example is the engineering of antibodies tailored to bind selectively to peptide antigens. \textbf{LaMBO2}~\cite{gruver2023protein} integrates BO and NOS to optimize the hu4D5 antibody for improved expression yield and binding affinity to the HER2 antigen. \textbf{TCR-dWAE}~\cite{li2023disentangled} is a disentangled Wasserstein autoencoder, can isolate the function-related patterns from the rest and is utilized for engineering T-cell receptors (TCRs) to bind specifically to peptide antigens.

\paragraph{Text.} Text-guided protein design allows practitioners to describe their demands in natural language without formal specifications. ProteinDT~\cite{liu2023text} is a pioneering method that leverages textual descriptions for protein design. \textbf{ProteinDT} consists of three modules: ProteinCLAP that aligns the representation of two modalities, a facilitator that generates the protein representation from the text modality, and a decoder that generates the protein sequences from the representation. \textbf{InstructProtein}~\cite{wang2023instructprotein} equips LLMs with the ability to comprehend protein language and further improve the performance of LLMs on protein design tasks through knowledge instruction tuning.

\section{Applications}
\label{sec:app}
\subsection{Peptide}
Peptides, characterized as short chains of amino acids with lengths varying from 2 to 50 units, exhibit a wide range of biological activities. Rather than simply screening an existing library, peptides with desired properties can be designed through generative models. Landmark studies include the design of signal peptides~\cite{wu2020signal} and antimicrobial peptides~\cite{das2021accelerated,szymczak2023discovering}.

\subsection{Antibody}
Antibodies are large, Y-shaped immunoproteins with both highly conserved parts and highly variable regions, formed by a heavy and a light chain, each comprising one variable domain and several constant domains. The variable domain is partitioned into a framework region and three complementarity determining regions (CDRs). The diversity inherent in CDRs empowers antibodies to confer immunity by specifically targeting and binding to a wide spectrum of pathogens. Thus, the majority of antibody design methods focus on redesigning or infilling CDRs while keeping the remainder unchanged~\cite{gruver2023protein,chen2024xtrimopglm}.

\subsection{Enzyme}
Enzymes, which are biological catalysts made primarily of proteins, may greatly accelerate the rate of chemical reactions. These substances are essential for life, performing numerous critical roles within the body, such as aiding in digestion and metabolism. For enzyme design, one may either locally modify an existing enzyme to enhance its activity~\cite{wang2023self} or synthesize diverse enzymes from scratch, thereby expanding the functional enzyme sequence space~\cite{repecka2021expanding,madani2023large}.

\section{Challenges and Opportunities}
\label{sec:future}
In this section, we discuss the open challenges that we face in protein sequence design in terms of data, algorithms, and evaluation, and identify potential opportunities.

\subsection{Challenges}
\begin{itemize}
    \item \textbf{Lack of assay-labeled data}: 
    The acquisition of assay-labeled data poses considerable challenges and incurs high costs, stemming from the complexity of \textit{in vitro} experimental procedures. For regimes where such labeled data are scarce, supervised predictive models are not accurate enough to guide the optimization process.
    \item \textbf{Lack of interpretability}: Owing to the `black box' nature of deep learning models, which serve as the foundation for current protein sequence design approaches, there is a notable interpretability challenge. Consequently, practitioners struggle to elucidate the critical functional sites involved in the protein generation or optimization processes, nor can they provide effective remedies for failures encountered during \textit{in silico} experiments.
    \item \textbf{Lack of reliable evaluation protocols}: Currently, the evaluation of protein sequence design models often relies on distribution-based computational metrics that may lack biological relevance and robustness. The development of a suite of task-specific, biologically meaningful, and robust evaluation metrics to identify promising protein candidates persists as an unresolved challenge.
\end{itemize}

\subsection{Opportunities}
\begin{itemize}
    \item \textbf{Build comprehensive benchmarks}: Scientific progress can be substantially accelerated if comprehensive and universally recognized benchmarks are established. We should therefore also strive for more reliable and comparable \textit{in silico} benchmarks. Within the realm of protein sequence design, researchers have recently established the first benchmark for protein inverse folding, ProteinInvBench~\cite{gao2023proteininvbench}. However, there remains an urgent need to create more realistic benchmarks to ensure equitable comparisons among various models tailored to the expansive spectrum of protein design challenges.  
    \item \textbf{Leverage PLMs}: RL is a prevalent methodology for the task-directed tuning of LLMs, with ChatGPT exemplifying its success. The integration of RL with PLMs for controllable design is nascent, yet holds immense promise. As PLMs gain prominence, there emerges a compelling opportunity to leverage RL in steering these pre-trained models to discover sequences that manifest predefined characteristics, while preserving naturalness.
    \item \textbf{Explore more generative frameworks}: Considering the evolutionary origins of proteins, it is insightful to explore the effectiveness of diverse non-autoregressive generative modeling approaches beyond diffusion models. Innovative frameworks including GFlowNet and Schrödinger Bridge present viable alternatives that merit further investigation within this scientific field.
\end{itemize}

\section{Conclusion}
In this paper, we systematically review the landscape of generative AI technologies and their recent progress in controllable protein sequence design. We present typical tasks, principal methodologies, and promising future directions in the field. We anticipate that this survey will provide a clear picture of this emerging field, encourage cross-disciplinary partnerships, and set a roadmap for both researchers and practitioners to move forward.

\appendix




\bibliographystyle{named}
\bibliography{ijcai24}

\end{document}